\documentclass[usenatbib,conference]{basi}

\usepackage[british]{babel}
\usepackage[varg]{txfonts}
\usepackage{rotating}
\usepackage{dcolumn}

\def\pa{\partial}
\def\vf{{\bf v}}

\begin{document}
\title[Solar Dynamo]{Dynamo Models of the Solar Cycle: Current Trends and Future Prospects}
\author[Dibyendu Nandy]%
       {Dibyendu Nandy\thanks{email: \texttt{dnandi@iiserkol.ac.in}}\\
       Department of Physical Sciences, Indian Institute of Science Education and Research, Kolkata,\\ Mohanpur 741252, West Bengal, India}

\pubyear{2011}
\volume{00}
\pagerange{\pageref{firstpage}--\pageref{lastpage}}

\date{Received \today}

\maketitle

\label{firstpage}

\begin{abstract}

The magnetic cycle of the Sun, as manifested in the cyclic appearance of sunspots, significantly influences our space environment and space-based technologies by generating what is now termed as space weather. Long-term variation in the Sun's magnetic output also influences planetary atmospheres and climate through modulation of solar irradiance. Here, I summarize the current state of understanding of this magnetic cycle, highlighting important observational constraints, detailing the kinematic dynamo modeling approach and commenting on future prospects.

\end{abstract}


\section{Introduction: The Solar Cycle}

\subsection{Scope of this Review}

The Sun is the most dynamically important astronomical body in the solar system. It's dynamism is primarily governed by its variable magnetic output. The magnetic field of the Sun produces sunspots and contributes to a range of physical processes ranging in timescales from minutes to centuries. Magnetic storms such as flares and Coronal Mass Ejections (CMEs) originate due to the sudden destabilization or restructuring of solar magnetic fields and affect the heliospheric environment, generating space weather. Slow, long-term changes in the Sun's magnetic and radiative output occur over longer timescales affecting planetary atmospheres and climate.

This review focuses on the mechanism that produces solar magnetic fields -- i.e., the solar dynamo mechanism and will not attempt to describe the plethora of activity that the generated magnetic field spawns. It is also a good idea to declare at the outset that this is not a comprehensive review of the subject and that the discussions and papers highlighted here are perhaps coloured by my personal perceptions. For a thoroughly comprehensive and freely available review on the subject of dynamo theory, interested readers are referred to Charbonneau (2005). This review is structured as follows: I summarize what I deem to be the most important constraints set by solar cycle observations on theories of the solar cycle (Section ~1.2), provide a concise description of the dynamo process (Section~2), introduce readers to kinematic dynamo models (Section~3) and finally discuss current trends and future prospects (Section~4).

\subsection{Observations of Solar Magnetic Fields and Plasma Flows}

\begin{figure}[t]
\centerline{\includegraphics[width=5.0cm]{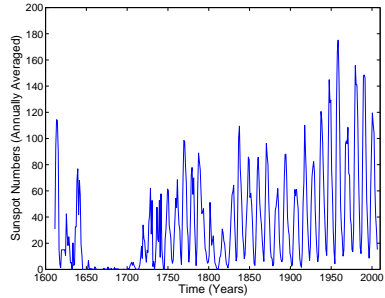}}
\caption{Sunspot observations over the last four centuries starting with the pioneering telescopic observations of Galileo. The 11 year periodic variation in the number of sunspots observed on the Sun's surface is clearly apparent. The observations also indicate a significant fluctuation in the amplitude of the cycle. The Maunder minimum, lasting between 1645--1715 AD, characterized a period over which very few sunspots were observed.}
\end{figure}

Dark spots on the Sun known as sunspots have been observed for centuries. The number of sunspots on the solar surface varies periodically with an average period of 11 years generating what is known as the solar cycle (Fig.~1). Following the discovery of strong magnetic fields -- on the order of $1000$ Gauss (G) -- within sunspots, it became obvious that sunspots point to the existence of a magnetic cycle in the Sun. These long-term sunspot observations provide the most important constraint on theories of the solar magnetic cycle. While the average period of the sunspot cycle is 11 years, both the period as well as the amplitude exhibit fluctuations. The amplitude of the cycle, in particular, varies significantly and in the recorded history, there is one episode between 1645 and 1715 AD known as the Maunder minimum (Eddy 1976), when almost no sunspots were observed on the Sun. At the beginning of a typical solar cycle, sunspots first appear at mid-latitudes in both the hemispheres. With the progress of the cycle, more and more spots appear at lower and lower latitudes, this equatorward migration generating the so called solar butterfly diagram. The cycle ends after 11 years when the sunspots are close to the equator. Subsequently, a new cycle begins with sunspots emerging again at higher latitudes. However, the relative orientation of the bipolar sunspot pairs reverse from one 11 year cycle to the next and therefore on considering the sign, the full magnetic cycle is of 22 years. The bipolar sunspot pairs also exhibit a systematic tilt with respect to the local parallel of latitude; this tilt angle is larger for sunspots emerging at higher latitudes (Hale et al.\ 1919), and follows an approximate cosine dependence on latitude (Joy's law).

There is a weak, small-scale magnetic field outside of sunspots which is observed to migrate polewards with the progress of the cycle (Bumba \& Howard 1965). This field usually exists in unipolar patches -- especially at high latitudes -- and is responsible for the cancelation and subsequent replenishment of the polar field of the Sun. The polar field reaches its maximum when the sunspot cycle is at its minimum and vice-versa. However, the cycle of the polar field has the same period as that of the sunspot cycle, indicating that they are connected to each other.

While it has been known for some time that the Sun rotates differentially with the equator rotating faster than the poles, with the advent of helioseismology, it has become possible to infer the internal differential rotation of the Sun (Charbonneau et al.\ 1999). In the solar convection zone (SCZ) -- the outer convective region of the Sun -- the rotation predominantly varies with latitude. At the base of the SCZ, within a region termed as the tachocline, the rotation varies with radius very rapidly. Deeper down, where helioseismic inferences are problematic, the radiative interior of the Sun is believed to rotate almost uniformly with no differential rotation. An axisymmetric, North-South flow of material is also observed in the surface (Hathaway 1996) which flows from the equator towards the poles. Helioseismic observations indicate that this flow pervades a major fraction of the SCZ (Giles et al.\ 1997); theoretical considerations such as mass conservation requires an equatorward counterflow deeper down. This flux of material is known as the meridional circulation, the peak (polweard) speed of which near the surface is about $20$ m/s, with significant intra- and inter-cycle fluctuations. Both differential rotation and meridional circulation play important roles in the generation and evolution of solar magnetic fields as we shall describe in the next section.

\section{The Dynamo Process}

Matter exists in the plasma state in the solar interior and the dynamics of magnetic fields in such a plasma system is governed by the magnetohydrodynamic (MHD) induction equation
\begin{equation}
\frac{\partial \bf{B}}{\partial t} =  {\nabla} \times (\bf{v} \times \bf{B} - \lambda \, \nabla
\times \bf{B)},
\end{equation}
where {\bf{B}} is the magnetic field, {\bf{v}} is the (plasma) velocity field and $\lambda$ is the magnetic diffusivity of the system. Astrophysical plasma systems have a high characteristic magnetic Reynolds number (the ratio of the first to the second term on the R.H.S.\ of the above equation), which implies that magnetic fields remain frozen in the plasma flows and the energy of plasma motions is coupled to that of the field. Therefore, the kinetic energy of plasma flows can be drawn in to generate magnetic fields. For example, if we assume that there is a large-scale North-South component of the magnetic field in the $r-\theta$ plane (the meridional plane) -- termed as the poloidal component, then this poloidal field will be stretched by the differential rotation of the Sun in the $\phi$-direction, generating a toroidal ($\phi$) component of the magnetic field. This toroidal component of the magnetic field is believed to be stored and amplified in a overshoot layer at the base of the SCZ where the temperature gradient is such that the fields are stable to buoyancy instabilities (Spiegel \& Weiss 1980, van Ballegooijen 1982). However, strong toroidal flux tubes can escape out of this stable layer due to overshooting convection and emerge into the SCZ -- where they become susceptible to magnetic buoyancy and rise up to the surface forming sunspots (Parker 1955a). The rising flux tubes are subject to the Coriolis force and pick up a tilt which is manifested in the tilt angle distribution of bipolar sunspot pairs.

For the dynamo cycle to be completed, the poloidal component of the field has to be regenerated back from the toroidal component. One of the first ideas proposed as a regeneration mechanism of the poloidal component is the dynamo $\alpha$-effect -- wherein, rising toroidal flux tubes are twisted back into the $r-\theta$ plane by helical turbulent convection (Parker 1955b). Turbulent convection can twist only those fields whose energy is roughly in equipartition with the energy of convective flows. This equipartition field is estimated to be on the order of $10^4$ G in the SCZ. However, simulations of the buoyant rise of toroidal flux tubes indicate that the initial magnetic field strength of sunspot-forming toroidal flux tubes is on the order of $10^5$ G at the base of the SCZ (Choudhuri \& Gilman 1987, D'Silva \& Choudhuri 1993, Fan, Fisher \& DeLuca 1993). Such strong flux tubes would quench the classical dynamo $\alpha$-effect and therefore an alternative mechanism is required for regenerating the poloidal field from super-equipartition flux tubes.

The alternative mechanism harks back to the ideas of Babcock (1961) and Leighton (1969), who proposed that the decay and dispersal of the flux of tilted bipolar sunspot pairs, mediated via turbulent diffusion, differential rotation and meridional flow can regenerate the poloidal field. Simulations of magnetic flux transport in the solar surface confirm that this process works (Wang, Sheeley \& Nash 1991). Recent observations linking the surface poloidal field source (tilted bipolar sunspot pairs) of a given cycle to the amplitude of the next cycle (Dasi-Espuig et al. 2010) also lends strong support to the Babcock-Leighton (BL) mechanism for poloidal field creation. In this process, the magnetic flux of the leading polarity spots from the two hemisphere cancel each other across the equator, while the magnetic flux of the following polarity spots preferentially move to high latitudes canceling the old cycle polar field and imparting the polarity of the current cycle. The poloidal component of the field thus generated is carried down into the interior through the combined action of turbulent diffusion, meridional circulation and downward flux pumping, where, it is stretched by the differential rotation to create the toroidal field of the next cycle.

\section{Kinematic Dynamo Models}

In dynamo parlance, the word ``kinematic'' is used to signify that the magnetic fields do not feed-back and change the underlying characteristics of the plasma flows; the latter therefore are not dynamic variables in the kinematic approach.
This approach to dynamo modeling attempts to capture the essential physics of the solar magnetic cycle (outlined in the previous section) by solving the MHD induction equation (Eqn.~1) in a solar-like geometry with prescribed plasma flow parameters and a poloidal field source. These parameters include differential rotation, meridional circulation and turbulent diffusion (see Mu\~noz-Jaramillo, Nandy \& Martens 2009 for a detailed treatment of these input parameters). Typically one invokes the assumption of axisymmetry and expresses the mean magnetic and velocity fields in terms of its toroidal and poloidal (meridional) components:
\begin{equation}
{\bf B} = B {\bf \hat{e}}_{\phi} + {\bf {B_p}}
\end{equation}
\begin{equation}
{\bf v} = r\sin(\theta)\Omega{\bf \hat{e}}_{\phi} + {\bf v_p}.
\end{equation}
The first and second term on the R.H.S.\ of Eqn.~$2$ are the toroidal and poloidal ($ {\bf B_p} = \nabla \times A {\bf \hat{e}}_{\phi}$) components of the magnetic field. In the case of the velocity field (Eqn.~3), these two terms correspond to the differential rotation $\Omega$ and meridional circulation ${\bf v_p} = {v_r}{\bf \hat{e}}_{r} + {v_\theta}{\bf \hat{e}}_{\theta}$. The diffusion coefficient in the induction equation (Eqn.~1) is replaced by the effective diffusivity $\eta$ -- which includes a (relatively larger) contribution from turbulent diffusivity -- for which an appropriate form is prescribed. In earlier models, a constant diffusivity used to be assumed. However, in recent and more realistic models, a depth-dependent diffusivity is prescribed -- which takes into account the fact that the SCZ is highly turbulent but the overshoot layer and the radiative zone beneath are relatively less so.

Assuming that the mean-field $\alpha$-effect does not contribute significantly to toroidal field generation, one can derive the $\alpha$$\Omega$ dynamo equations by substituting Eqns.~2 and 3 into Eqn.~1 and separating out the the toroidal and poloidal components:
\begin{eqnarray}
\frac{\pa B}{\pa t} + \frac{1}{r} \left[ \frac{\pa}{\pa r} (r v_r B) + \frac{\pa}{\pa
\theta}(v_{\theta} B) \right] + \nabla \eta \times (\nabla \times B) = \eta \left( \nabla^2 -
\frac{1}{s^2} \right) B + s({\bf B_p} \cdot \nabla)\Omega,
\end{eqnarray}
\begin{eqnarray}
\frac{\pa A}{\pa t} + \frac{1}{s}(\vf_p \cdot \nabla)(s A) = \eta \left( \nabla^2 - \frac{1}{s^2}
\right) A + {S_{p}},
\end{eqnarray}
where $s = r sin\theta$. The toroidal and poloidal field evolution are described by Eqn.~4 and Eqn.~5 respectively. Note that the magnetic and velocity fields in the induction equation can be expressed as a sum of a mean and a fluctuating component. The field descriptors appearing in the toroidal and poloidal field evolution equations correspond to the mean fields.

The last term in Eqn.~5, $S_p$, denotes the added source term responsible for the regeneration of the poloidal component. This could be a prescribed mean-field $\alpha$-effect (resulting out of cross-correlations between the fluctuating velocity and magnetic field components) of the form $S_p = \alpha B$, often incorporating an amplitude-limiting quenching factor that signifies the non-linear feedback of the magnetic fields on the flows, included through a parametrization of the $\alpha$-effect such as $\alpha \sim \alpha_0/[1 + (B/B_{quenching})^2]$, where $\alpha_0$ is the amplitude of the $\alpha$-effect and $B_{quenching}$ is the field strength at which non-linear feedback becomes effective. Note that this parametrization is still within the domain of a kinematic approach as the mean flows themselves are not dynamical variables.
In most models, the Babcock-Leighton mechanism of poloidal field creation is captured indirectly, by adopting a functional form of the $\alpha$-effect which limits it to near-surface layers, and by making the source proportional to the toroidal field at the base of the base of the SCZ (i.e., a non-local $\alpha$-effect formulation; Dikpati \& Charbonneau 1999). A more sophisticated treatment of the Babcock-Leighton type dynamo source explicitly erupts toroidal fields exceeding a buoyancy threshold to the near-surface layers where the $\alpha$-effect works on it (Nandy \& Choudhuri 2001) or implements a double-ring algorithm that realistically captures the surface flux-transport mechanism leading to polar field reversal and creation (Mu\~noz-Jaramillo et al.\ 2010). Note that in spirit, the Babcock-Leighton $\alpha$-effect is not a mean-field $\alpha$-effect in the sense that it does not involve averaging over small-scale fluctuations or the first order smoothing approximation (for further details, see Charbonneau 2005 and Nandy 2010a).

The two equations (Eqns.~4 and 5) are solved with the specification of boundary conditions appropriate for the Sun. Fig.~2 shows a simulated butterfly diagram from such a kinematic solar dynamo model with a non-local Babcock-Leighton poloidal field source. Note that alternative formulations of the kinematic dynamo models exists in which the dynamo $\alpha$-effect owes its origins to other mechanisms, such as the traditional mean-field $\alpha$-effect due to helical turbulence in the main body of the SCZ, or the buoyancy instability of toroidal flux tubes near the base of the SCZ. Two or more poloidal field sources can also in principle be prescribed to act simultaneously, but at different layers and at different operating thresholds. A detailed description of all these different formulations is beyond the scope of this review, for which interested readers are referred to Charbonneau (2005). The point to take home is that kinematic dynamo modeling is fairly versatile and can capture a diverse range of physics and phenomena, with properly constrained and well-thought out simulation runs.

\section{Current Trends and Future Prospects}

Kinematic dynamo simulations have made major contributions to our understanding of the solar cycle in the last decade or so. To begin with, it is now clear that the meridional flow of plasma, especially the equatorward counterflow in the meridional circulation plays an important role in the equatorward migration of the sunspot zone and in setting the period of the sunspot cycle (Choudhuri, Sch\"ussler \& Dikpati 1995; Durney 1995; Dikpati \& Charbonneau 1999; Nandy \& Choudhuri 2002; Yeates, Nandy \& Mackay 2008). The equatorward meridional counterflow achieves this by transporting the poloidal field of the previous cycle and the new toroidal field (that is being generated) very efficiently towards the equator through the weakly turbulent overshoot layer and upper parts of the radiative zone. However, the vertical transport of magnetic flux through the convection zone is shared to varying degrees between meridional circulation, turbulent diffusion and downward flux pumping -- an issue we will return to later.

\begin{figure}[t]
\centerline{\includegraphics[width=5.0cm]{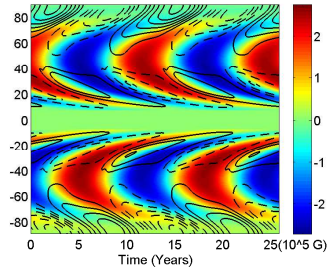}}
\caption{A time-latitude plot of the toroidal field at the base of the solar convection zone (in colour) and the surface radial field (contours) from a kinematic solar dynamo simulation by Mu\~noz-Jaramillo, Nandy \& Martens (2009). Positive radial field is indicated by solid contours and negative radial field by dashed contours. This simulated butterfly diagram shows that such dynamo models are able to match the important observational features of the solar cycle. Cyclic reversal of the toroidal field, from which sunspots are generated is clearly seen, in conjunction with the evolution of the surface radial field -- which contributes to polar field reversal.}
\end{figure}

The amplitude of the cycle is known to be modulated by various mechanisms. The primary constraint is set by magnetic buoyancy, which determines the lower operating threshold for the poloidal field source (Nandy 2002). Non-linear feedbacks due to the back-reaction of the magnetic field on the flows could also be playing a role here. In the context of the BL mechanism, one such process for limiting cycle amplitude is discussed by Cameron (this volume). Small fluctuations in the amplitude of the solar cycle can be induced by the stochastic nature of the dynamo $\alpha$-effect (Charbonneau \& Dikpati 2000), including the scatter in the active region tilt angle distribution, or through fluctuations in the meridional plasma flow (Karak \& Choudhuri 2011). Meridional flow variations are also responsible for modulation of the solar cycle period and the length and nature of solar cycle minima (Nandy, Mu\~noz-Jaramillo \& Martens 2011). While kinematic dynamo simulations indicate that meridional circulations plays a vital role in various aspects of solar activity, a matter of continued frustration is our inability to probe the deeper flow (including the equatorward counterflow) in the SCZ -- where much of the mass flux in the circulation lies.

There is, however, no consensus opinion on the cause of global minima such as the Maunder minima in solar activity -- periods when hardly any sunspots are observed. Long-term reconstructions of solar activity based on cosmogenic isotopes indicate that such global minima have also occurred in the past -- from which the sunspot cycle recovers to normal levels of activity. While some attribute the cause to non-linear feedback mechanisms (Wilmot-Smith et al.\ 2005 and references therein), some attribute this to stochastic fluctuations (Charbonneau, Blais-Laurier \& St-Jean 2005), and others to meridional flow variations (Karak 2010). Outstanding questions remain however as to how the cycle recovers from a deep minima with a prolonged period without the BL poloidal field source. Questions also remain as to what happens to the parity of the solar cycle when the dynamo enters a global minima and whether the same parity is preserved when the cycle recovers from such grand minima episodes.

The parity of the solar cycle sets important constraints on the dynamo mechanism itself and on the relative roles of various flux transport processes. The Sun is observed at present to exhibit the odd parity -- wherein the poloidal field is of dipolar nature and the toroidal field is of opposite sign (i.e., the relative orientation of the bipolar sunspots pairs are reversed) in the two hemispheres. This configuration is not necessarily always guaranteed. For example, some full-sphere dynamo simulations (encompassing both the hemispheres) show a drift or change in parity. This is especially true in situations where very low turbulent diffusivity is assumed in the SCZ (Dikpati \& Gilman 2001) which led the latter to argue that the BL dynamo violates the observed parity in the Sun. However, Chatterjee, Nandy \& Choudhuri (2004) has shown that elevated levels of turbulence -- as is appropriate in the main body of the SCZ -- is able to preserve the observed parity for BL dynamos by efficiently coupling the two hemispheres; this is confirmed by Hotta \& Yokoyama (2010).

The finite time delay in the transport of magnetic flux by flux transport processes also introduce a memory in the solar cycle -- which besides leading to interesting activity dynamics (Wilmot-Smith et al.\ 2006), also lays the basis for solar cycle predictions. Unfortunately, the two dynamo based solar cycle predictions to date have generated conflicting results, one of a very strong cycle (Dikpati, de Toma \& Gilman 2006) and one of a weak cycle (Choudhuri, Chatterjee \& Jiang 2007). The cause of this conflicting result has been identified in the relative roles attributed to  meridional circulation and turbulent diffusion (Yeates, Nandy \& Mackay 2008). While Dikpati et al.\ assume a low turbulent diffusivity, Choudhuri et al.\ assume a high turbulent diffusivity in the SCZ. It is not clear which of these assumptions is valid, although, independent studies (such as those addressing the parity issue) point out that turbulent diffusivity is likely to be high in the SCZ. The issue is further confounded by the fact that both the models used for cycle prediction ignored the downward pumping of magnetic flux by turbulent convection in a rotating system -- which magnetohydrodynamic simulations show to be an important flux transport process within the SCZ (Tobias et al.\ 2001).

Evidently some of the important advances in this subject in the near future will come from addressing the issues that are currently deemed as major problems (see e.g., Nandy 2010b). This will include (in no particular order) probing the structure of the deep meridional circulation -- including the equatorward counterflow, exploring the causes of deep minima in solar activity -- such as the Maunder minimum, and exploring the relative roles played by different flux transport processes -- especially turbulent flux pumping -- in the solar cycle. Obviously, quantitatively accurate predictions of the solar cycle will be dependent on the understanding gleaned from such studies.


A complete understanding of the solar cycle requires close synergy between three diverse modeling approaches. The focus of my discussion here was only on one of them -- namely kinematic dynamo simulations. The other two tools are full MHD simulations (where equations for the evolution of the flows and magnetic fields are solved self-consistently in conjunction with the energy equation) and solar surface flux transport simulations. Surface flux transport simulations have not only successfully demonstrated how the BL mechanism of poloidal field generation works, they also provide a natural bridge between the evolution of the magnetic field in the solar interior and that in the corona and heliosphere -- where the fields respond to variations on the Sun's surface. Kinematic dynamo models must therefore assimilate the best features of surface flux transport models towards addressing the issue of how the solar interior couples to the heliosphere. This goal looks achievable with dynamo models progressing to the stage where they can now realistically capture the essential characteristics of surface flux transport models (Mu\~noz-Jaramillo et al.\ 2010). Full MHD simulations of the solar interior provides a necessary and important complement to kinematic dynamo simulations. Much of the processes in the interior of the Sun is hidden from view and is not yet accessible to helioseismology; under the circumstances, MHD simulations reveal some of the important processes that should be included in kinematic dynamo models of the solar cycle (e.g., the downward pumping of magnetic flux). The full potential of MHD models of the solar interior seems to be within grasp with the recent, successful simulation of cyclic reversal of magnetic fields (Ghizaru, Charbonneau \& Smolarkiewicz 2010) -- a task that had long remained elusive. We seem to be well positioned for an era of discovery regarding the solar cycle.

\section*{Acknowledgements}

I am grateful to the Department of Science and Technology, Government of India for supporting my research through the Ramanujan Fellowship.

\label{lastpage}

\end{document}